\documentclass[preprintnumbers,prd,onecolumn,showpacs,floatfix,superscriptaddress, nofootinbib]{revtex4}
\usepackage{graphicx}
\usepackage{epsfig}
\usepackage{bm}
\usepackage{amssymb}
\usepackage{float}
\usepackage{amsmath}
\usepackage{subfigure}
\usepackage{dcolumn}
\usepackage{cancel}
\usepackage[colorlinks]{hyperref}
\usepackage[usenames,dvipsnames]{color}
\hypersetup{
     breaklinks=true,
    pdfstartview={FitH},    
    colorlinks=true,       
    linkcolor=blue,          
    citecolor=red,        
    filecolor=magenta,      
    urlcolor=blue,           
    anchorcolor=green,      
    linktocpage=true
}

\newcommand{\Mpl}{M_{\textrm{Pl}}}
\renewcommand{\(}{\left(}
\renewcommand{\)}{\right)}
\renewcommand{\k}{\vec{k}}
\newcommand{\nn}{\nonumber}

\def\Om{\Omega}
\def\lam{\lambda}
\def\doi{http://doi.org}
\def\arxiv{http://arxiv.org/abs}
 \def\t{\tilde}
\def\r{\mathrm{r}}
\def\g{\mathrm{g}}
\def\h{\mathrm{h}}
\def\m{\mathrm{m}}

\begin{document}

 \title{Relic  gravitational waves from Quintessential Inflation}

\author{Safia Ahmad}
\email{safia@ctp-jamia.res.in}
\affiliation{Centre for Theoretical Physics, Jamia Millia
Islamia,
New Delhi-110025, India}

\author{R. Myrzakulov}
\email{rmyrzakulov@gmail.com} \affiliation{ Eurasian  International
Center for Theoretical Physics, Eurasian National University, Astana
010008, Kazakhstan}
\author{M.~Sami}
\email{samijamia@gmail.com}
\affiliation{Centre
for Theoretical Physics, Jamia Millia Islamia, New Delhi-110025,
India}
\affiliation{ Eurasian  International
Center for Theoretical Physics, Eurasian National University, Astana
010008, Kazakhstan}
\affiliation{Maulana Azad National Urdu
University, Gachibowli, Hyderabad-500032,India} 

\begin{abstract}
We study relic gravitational waves in the paradigm of quintessential inflation. In this framework, irrespective of the underlying model, inflation is followed by the kinetic regime. Thereafter, the field energy density remains sub-dominant before the onset of acceleration. We carry out model independent analysis to  obtain the temperature at the end of inflation
and  the estimate for upper bound on the Hubble parameter to circumvent the problem
due to relic gravitational waves.
In this process, we used Planck 2015 data to constrain the inflationary phase. We  demonstrate that the required temperature can be produced by the
mechanism of instant preheating. The generic feature of
the scenario includes the presence of kinetic regime after inflation
which results into blue spectrum of gravitational wave background at
high frequencies. We discuss the prospects of detection of relic gravitational wave background in the
advanced LIGO and LISA space-born gravitational wave missions. Finally we consider a concrete model to realize the paradigm of quintessential inflation and show that inflationary as well as post-inflationary evolution can successfully be described by the  inflaton potential,  $V(\phi)\propto Exp(-\lambda \phi^n/\Mpl^n)(n>1)$, by suitably constraining the parameters of the model.
\end{abstract}


\maketitle

\section{Introduction} \label{sec:introduction}

Accelerated expansion has played an important role in the history of
our universe. It is a common belief  that Universe has gone through
a phase of rapid expansion dubbed {\it inflation}
\cite{Starobinsky:1980te,Starobinsky:1982ee,Guth:1980zm,Linde:1983gd} at early times and it started accelerating once again
around the present epoch \cite{Copeland:2006wr,Sahni:2006pa,Peebles:2002gy,Sami:2009dk,Sami:2009jx,Sami:2013ssa}. The standard model of Universe, therefore, should be
complemented by two phases of acceleration. Inflation is a beautiful
paradigm that not only resolves the inconsistencies of hot big bang
model but also provides with a mechanism of generation of primordial
perturbations required for structure formation. Late-time cosmic
acceleration is needed to address the age problem in the standard
model. The phenomenon was detected in 1998 by supernovae
observations \cite{Riess:1998cb,Perlmutter:1998np} which was supported by other probes
indirectly thereafter \cite{Bennett:2003bz,Lange:2000iq,Ade:2015tva}. Similar
confirmation is yet awaited for cosmological inflation.

Cosmic acceleration is a generic phenomenon in our Universe which
manifests at early and late times keeping the thermal history
intact. Often, these two phases are described independently. As for
late-time acceleration, a variety of scalar field models has been
investigated in the literature since 1998. At the background level,
 cosmology community seems to converge on the cosmological constant
as the underlying cause of the said phenomenon. Even if acceleration
is caused by a slowly rolling scalar field, the latter is not
distinguished from cosmological constant at the background level. As
for perturbations, the study of large scale structure might reveal
in case there is something beyond $\Lambda$CDM. The current
observational constraints related to inflation are tight such that
many popular models are on the verge of being ruled out. We should
 emphasize that in a successful model, inflation should be
followed by an efficient reheating also.

It is interesting to ask whether a successful model of inflation can
also describe late-time acceleration without interfering with the
thermal history of Universe or can inflaton be dark energy.
Unification of inflation with late-time acceleration or dark energy
is termed as {\it quintessential inflation}
\cite{Spokoiny:1993kt,Peebles:1999fz,Peloso:1999dm,Dimopoulos:2000md,Copeland:2000hn,Sahni:2001qp,Dimopoulos:2001ix,
Majumdar:2001mm,Giovannini:2003jw,Sami:2004xk,Sami:2004ic,Rosenfeld:2005mt,
Hossain:2014xha,Hossain:2014coa,Hossain:2014zma,Dimopoulos:2017zvq}.
It is indeed challenging to describe inflation and late-time
acceleration using a single scalar field. At the onset, it sounds
possible if the field potential is shallow at early and late stages
and steep for most of the history of Universe. First, the model
should comply with all the observational constraints related to
inflation and should give rise to successful reheating which  is
itself a difficult task. As we pointed out, the current
observational constraints are quite tight putting many known models
in tension. Secondly, conventional reheating mechanism
\cite{Dolgov:1982th,Abbott:1982hn,Ford:1986sy,Spokoiny:1993kt,Kofman:1994rk,Shtanov:1994ce,
 Kofman:1997yn,Campos:2002yk} is not
applicable to this class of models as the field potential is
typically run-away type. However, in this case, one can invoke an
alternative mechanism dubbed instant preheating
\cite{Felder:1998vq,Felder:1999pv,Campos:2004nc,Tashiro:2003qp}. It
is desirable that we have scaling regime \cite{Steinhardt:1999nw}
after inflation allowing inflaton to go into hiding till late times
which is necessary to keep the thermal history intact. To what
extent, the field should be hidden is decided by the nucleosynthesis
constraint. In this case, unlike the thawing picture, evolution, to great extent, is independent of initial conditions.

The generic feature of quintessential inflation includes the
presence of kinetic regime
\cite{Sahni:2001qp,Sami:2004xk,Tashiro:2003qp} that follows
inflation. The duration of the regime depends upon the
temperature at the end of inflation. The energy density of relic gravitational waves
~\cite{Grishchuk:1974ny,Starobinsky:1979ty,Sahni:1990tx,Souradeep:1992sm,Gasperini:1992dp,Gasperini:1992pa,Giovannini:1999bh,Giovannini:1999qj,Langlois:2000ns,
Tashiro:2003qp} as compared to field energy density enhances during kinetic regime and might challenge the
nucleosynthesis constraint at the commencement of radiative regime.
The distinguished signature of the unification, irrespective of the
underlying model, includes the production of relic gravitational
waves with blue spectrum. Clearly, an alternative reheating
mechanism is needed in this case to circumvent the
problem.


Let us note that the generalized exponential potential $V(\phi)\propto Exp(-\lambda
\phi^n/\Mpl^n),(n>1)$ \cite{Geng:2015fla} can successfully realize the paradigm of quintessential inflation. This potential has a remarkable property: it is
shallow around $\phi=0$ and steep thereafter. Thanks to an
additional parameter $n$ compared to standard exponential, the
model can comply with observational constraints on inflation.
Secondly, the model shares the property of a simple steep
exponential potential
\cite{Hossain:2014xha,Sahni:2001qp,Copeland:2000hn,Lucchin:1984yf,Ratra:1989uz,Sami:2002fs} at late
times, namely, the approximate scaling regime is a late-time attractor in this
case as $\Gamma=V'' V/V'^2\to 1$ for large values of the field.

Last but not least, we need to exit from scaling regime around the
present epoch. This can be achieved by invoking non-minimal coupling
to massive neutrino matter (see \cite{Wetterich:2013aca,Wetterich:2013jsa,Wetterich:2013wza,Wetterich:2014eaa,
Wetterich:2014bma,Amendola:2007yx,Wetterich:2007kr,Hossain:2014zma,Geng:2015fla}
and references therein for details). The coupling appears at late stages as neutrinos turn non-relativistic. As a
result, the effective potential for the field acquires minimum where
the field can settle causing exit from scaling regime to late-time
cosmic acceleration.

In this paper, we first carry out a model independent analysis to obtain the temperature at the end of inflation and the estimate on the upper bound of the Hubble parameter to circumvent the problem
due to relic gravitational waves.  We also investigate an alternative reheating mechanism suitable to the scenario under consideration. Finally, we discuss a model of quintessential inflation which can successfully realize the paradigm. Throughout this paper, $\Mpl$ denotes the reduced Planck mass.


\section{Relic gravitational wave spectrum}
\label{RGW}
In this section, we shall study relic gravitational waves in the
scenario of quintessential inflation. The
evolution of tensor perturbations, $h_{ij}$, is given by the
Klein-Gordon equation
\begin{equation}
\Box h_{ij}=0 \to \ddot{\varphi}_k(\tau)+2\frac{\dot
{a}}{a}\varphi_k(\tau)+k^2\varphi_k(\tau)=0~;~~h_{ij}\sim
\varphi_ke^{ikx}e_{ij} \, ,  \label{eq:KG}
\end{equation}
where $e_{ij}$ is the polarization tensor, $\tau$ (${\rm d}\tau={\rm
d}t/a$) is the conformal time and $k$ is the comoving wave number.
For simplicity, we shall  consider the exponential inflation ,
namely, $a=\tau_0/\tau$ and $H_{\rm in}=-1/\tau_0$.  The positive
frequency solution of Eq.~(\ref{eq:KG}) in the adiabatic vacuum,
corresponds to the ``in" state, $\varphi_{\rm in}^{(+)}(k,\tau)$
\begin{equation}
 \varphi_{\rm in}^{(+)}(k,\tau)=(\pi \tau_0/4)^{1/2}(\tau/\tau_0)^{3/2}H^{(2)}_{3/2}(k\tau)\,.
\end{equation}
where $H^{(2)}_{3/2}$ is the Hankel function of second kind.

In a scenario of quintessential inflation, Universe enters into the
kinetic phase with stiff equation of state parameter soon after the
inflation ends\cite{Sahni:2001qp,Sami:2004xk}. This transition
involves the non-adiabatic change of space-time geometry. Assuming
for simplicity that the post-inflationary dynamics is described by
the power law expansion, {\it i.e.}, $a=(t/t_0)^p\equiv
(\tau/\tau_0)^{1/2-\mu}$ where $ \mu\equiv (3(w-1)/2(3w+1))$ with
$w$ being the post-inflationary equation of state parameter. Let us note that
$\mu=0$ in the kinetic regime ($w=1$). As for the  ``out'' state, it
contains both positive and negative frequency solutions of
(\ref{eq:KG}),
\begin{equation}
\varphi_{\rm out}=\alpha \varphi_{\rm out}^{(+)}+\beta \varphi_{\rm
out}^{(-)} \, ,
\end{equation}
where $\alpha$ and $\beta$ designate the Bogoliubov coefficients
\cite{Sahni:2001qp} and
\begin{equation}
 \varphi^{(+,-)}_{\rm out} =(\pi \tau_0/4)^{1/2}(\tau/\tau_0)^\mu H^{(2,1)}_{|\mu|}(k\tau) \, .
\end{equation}
We then estimate the energy density of relic gravitational waves
\cite{Sahni:2001qp,Sahni:1990tx},
\begin{equation}
\label{rhog}
 \rho_g=<T_{00}>=\frac{1}{\pi^2a^2}\int{{\rm d}kk^3|\beta|^2} \, .
\end{equation}
During kinetic regime, $|\beta_{\rm kin}|^2\sim (k\tau_{\rm
kin})^{-3}$, using then Eq.~(\ref{rhog}) and  the fact that $H_{\rm
in}=-1/\tau_0$, we obtain,
\begin{equation}
\label{rhog2}
\rho_g=\frac{32}{3\pi}h^2_{\rm
GW}\rho_b\left(\frac{\tau}{\tau_{\rm kin}}\right)
\end{equation}
where $\rho_b$ denotes the background energy density and contains
radiation and stiff scalar matter in the kinetic regime. We have
hereby  assumed that the generation of radiation took place at the end of inflation
thank to some mechanism to be specified in the subsequent section.

Since, at the equality of radiation and scalar field energy
densities ($\tau=\tau_{\rm rh}$),
\begin{equation}
\label{taueq}
 \frac{\tau_{\rm rh}}{\tau_{\rm kin}}=\(\frac{T_{\rm kin}}{T_{\rm rh}}\)^2
\end{equation}
and $\rho_b=2 \rho_{r}$, we find using Eq.~(\ref{rhog2}),
\begin{equation}
\label{rhogr} \left(\frac{\rho_g}{\rho_{r}}\right)_{\rm
rh}=\frac{64}{3\pi}h^2_{\rm GW}\left(\frac{T_{\rm end}}{T_{\rm
rh}}\right)^2 \, ,
\end{equation}
where $h_{\rm GW}$ is the dimensionless amplitude of the
gravitational waves given by,
\begin{eqnarray}
\label{hgw}
 h^2_{\rm GW}=\frac{H^2_{\rm inf}}{8 \pi \Mpl^2}=\frac{V_{\rm {inf}}}{24 \pi \Mpl^4}\, ,
\end{eqnarray}
where $V_{\rm {inf}}$ is the value of the inflationary potential at the
time the cosmological scales exit the horizon and can be fixed by imposing the COBE
normalization \cite{Bunn:1996da,Bunn:1996py}.

Eqs.~(\ref{taueq}) and (\ref{rhogr}) implies that longer kinetic regime, {\it i.e.},
large $\tau_{\rm rh}$, would corresponds to smaller $T_{\rm rh}$ and hence the larger value of $\rho_g / \rho_{r}$ at equality. As given in \cite{Sahni:2001qp}, it can be shown from (\ref{rhog}) that for $w>1/3$, $\rho_g \propto 1/a^4$ and for
$w<1/3$, $\rho_g\propto \rho_b$; during radiation era also $\rho_g$ approximately
tracks the background. For $w=1/3$ (radiation), log factor appears \footnote{Precisely, $\rho_g \propto a^{-4} \log(\tau/\tau_0) \propto a^{-4} \log{(a)}$ but we have ignored the logarithmic factor as its contribution is negligible during evolution.}.

For simplicity \footnote{It is supported by numerical simulation.}, we hereby assume that kinetic regime ($\rho_\phi \propto 1/a^6$) immediately follows inflation, thus we can use the approximation, $H_{\rm end}\simeq H_{\rm
kin}$ and $T_{\rm end}\simeq T_{\rm kin}$. Since $T_{\rm rh}\sim
T_{\rm end} (a_{\rm end}/a_{\rm rh})$, we find,
\begin{equation}
\left(\frac{\rho_\phi}{\rho_{r}}\right)_{\rm end}=\left(\frac{T_{\rm
end}}{T_{\rm rh}}\right)^2
\end{equation}
and the following important relation
\begin{equation}
\left(\frac{\rho_\phi}{\rho_{r}}\right)_{\rm
end}=\frac{3\pi}{64}\left(\frac{\rho_g}{\rho_{r}}\right)_{\rm
rh}\frac{1}{h^{2}_{\rm GW}} \, . \label{eq:rhog_rhor}
\end{equation}

From COBE normalisation, we find that 
\begin{equation}
\label{vhinf}
V_{\rm {inf}}^{1/4}=0.013 r^{1/4} \Mpl,
\end{equation}
where $r$ is the scalar-to-tensor ratio. Combining this with Eq.~(\ref{hgw}) and inserting into the Eq.~(\ref{eq:rhog_rhor}), we find
\begin{equation}
\(\frac{\rho_{\phi}}{\rho_r} \)_{\rm {end}}=
\frac{9\pi^2}{8}\left(\frac{\rho_g}{\rho_r}\right)_{\rm rh}
\frac{M_{\rm Pl}^4}{V_{\rm inf}} \lesssim \frac{3.88 \times 10^6}{r}
\label{eq:rhog_r_bound}
\end{equation}
where we have used the constraint on the ratio, $(\rho_g/\rho_r)_{\rm rh}\lesssim 0.01$ as imposed by nucleosynthesis. We can compute temperature at the end of inflation by using the fact that the cosmological scales, $k$ that crosses the horizon during inflation and the scales that re-enters it at a later time satisfies the relation, $k=a_{\rm inf} H_{\rm inf}=a_0 H_0$, which implies that
\begin{eqnarray}
\label{hor_cross}
\frac{k}{a_0 H_0}=\frac{a_{\rm inf}}{a_{\rm end}}\frac{a_{\rm end}}{a_0}\frac{H_{\rm inf}}{H_0} \Rightarrow e^{-\mathcal{N}}\frac{T_0}{T_{\rm end}}\frac{H_{\rm inf}}{H_0}=1.
\end{eqnarray}
For $\mathcal{N}=65$, we find that
\begin{equation}
\label{T_end}
T_{\rm end}=1.99 \times 10^{15} r^{1/2}~\text{GeV}.
\end{equation}
Now, the upper bound on the value of Hubble parameter at the end of inflation, $H_{\rm end}$, can be obtained from Eq.~(\ref{eq:rhog_r_bound}) and using the fact that $\rho_{\rm {r,end}} = T_{\rm end}^4$,
\begin{equation}
\label{bound}
H_{\rm end} \lesssim 1.85 \times 10^{15} r^{1/2}~\text{GeV}.
\end{equation}
However, the bound (\ref{bound}) should be consistent with the estimate for $H_{\rm inf}$ obtained after using COBE normalization, namely, $H_{\rm inf}\simeq 2.37 \times 10^{14} r^{1/2}$GeV. In general, the ratio $H_{\rm inf}/H_{\rm end}$ is more than few, in the realistic scenarios with graceful exit. Clearly, $H_{\rm inf}=H_{\rm end}$ in case of  de-Sitter. Thus we can take $H_{\rm end} \lesssim H_{\rm inf}$ for estimates which is clearly a more conservative bound than (\ref{bound}). The bound on $H_{\rm end}$ in turn gives us the bound on $V_{\rm end}$ which is $V_{\rm end} \lesssim 1.15 \times 10^{-6}r~ \Mpl^4$. Again, we know that $V_{\rm end}$ can at most take the value of $V_{\rm inf}$.

Since the potential in the model of quintessential inflation is generically run-away type, the conventional reheating mechanism does not operate
in this case. Gravitational particle production is a known possibility. After inflation, the geometry of space-time undergoes a
non-adiabatic process which gives rise to particle production.
Let us note that gravitational particle production gives rise to the energy density \cite{Ford:1986sy,Spokoiny:1993kt}
\begin{equation}
\rho_{\rm {r,end}} \simeq 0.01 g_p H_{\rm end}^4
\end{equation}
where $g_p$ is the number of different particles produced which is typically $\sim 10-100$ and since $H_{\rm end} \lesssim H_{\rm inf}$, we find that
\begin{eqnarray}
\label{Spokoiny}
\left(\frac{\rho_{\phi}}{\rho_{r}}\right)_{\rm {end}} \simeq \frac{\rho_{\rm \phi,end}}{0.01 g_p H_{\rm end}^4} \gtrsim \frac{2.1 \times 10^{10} g_p^{-1}}{r}.
\end{eqnarray}
In the aforesaid model independent discussion, we have used $H_{\rm end}\simeq H_{\rm inf}$. For an exact  relation between these quantities, we need a concrete model. However, in models with $r \ll 1$ such as Starobinsky inflation model \cite{Starobinsky:1980te},  $H_{\rm inf}$ is close to $H_{\rm end}$. Thus, in general, the respective energy density produced due to gravitational particle production misses the nucleosynthesis constraint imposed by relic gravitational waves (\ref{eq:rhog_r_bound}) by at least two orders of magnitude which is related to the fact that temperature of radiation due to gravitational particle production $T_{\rm end}^g $ is some what less than $T_{\rm end}$ (\ref{T_end}), namely, $T_{\rm end}^g\simeq  2.63\times 10^{14} r^{1/2}$GeV.
In the scenario under consideration, due to the steep post-inflationary
behavior of the potential, the field enters the kinetic regime soon
after inflation ends and remains there for a long time in view of the aforesaid. Clearly, the problem
of relic gravitational waves in quintessential inflation is
associated with the longevity of kinetic regime. During this time
the energy density in gravitational waves enhances compared to field
energy density challenging  the nucleosynthesis
constraint at the commencement of radiative regime. We naturally
require more efficient process  to circumvent the problem,  one
should then look for a suitable reheating mechanism.

We are now in position to discuss the gravitational wave spectrum. The fractional spectral energy density parameter
of the relic gravitational wave is defined as follows,
\begin{equation}
 \Omega_{\rm GW}(k)=\frac{\t\rho_\g(k)}{\rho_{\rm c}} \, ,
\end{equation}
where $\rho_{\rm c}$ denotes the critical energy density (see Ref.
\cite{Sahni:2001qp} for details) and $\tilde{\rho}_\g(k)$ is the
spectral energy density
\begin{equation}
\tilde{\rho}_\g(k)\propto
k^{1-2|\mu|};~~\mu=\frac{3}{2}\left(\frac{w-1}{3w+1}  \right).
\end{equation}
From the above equation, we see that during the kinetic regime, {\it
i.e.}, for $w=1$, $\rho_\g(k)\propto k$ giving rise to a blue spectrum
in the  wave spectrum. This is a generic feature of any
 model of quintessential inflation where the scalar field enters into the kinetic regime after inflation.

We also have the following relations,
\begin{eqnarray}
&&\Omega_{\rm GW}^{\rm (MD)}= \frac{3}{8\pi^3}h_{\rm GW}^2
\Omega_{\m 0}\(\frac{\lam}{\lam_\h}\)^2 \, ,  \lam_{\rm MD}<\lam\leq \lam_\h  ,~~~~~~\\
\label{rom}
&&\Om_{\rm GW}^{\rm (RD)}(\lam)=\frac{1}{6\pi}h_{\rm GW}^2\Omega_{\r
0} \, ,
~~~~~~~~~ \lam_{\rm RD}<\lam\leq\lam_{\rm MD} \, ,~~~~~ \\
&&\Om_{\rm GW}^{\rm (kin)}(\lam)=\Om_{\rm GW}^{\rm
(RD)}\(\frac{\lam_{\rm RD}}{\lam}\) \, , ~~~~~  \lam_{\rm
kin}<\lam\leq\lam_{\rm RD} \, ,~~~~~
\end{eqnarray}
with
\begin{eqnarray}
 \lam_\h &=& 2cH_0^{-1}\, , \\
 \lam_{\rm MD} &=& \frac{2\pi}{3}\lam_\h\(\frac{\Om_{\r 0}}{\Om_{\m 0}}\)^{1/2} \, , \\
 \lam_{\rm RD}&=& 4\lam_\h\(\frac{\Om_{\m 0}}{\Om_{\r 0}}\)^{1/2}\frac{T_{\rm MD}}{T_{\rm end}} \, ,\\
 \lam_{\rm kin} &=& cH_{\rm kin}^{-1}\(\frac{T_{\rm end}}{T_0}\)\(\frac{H_{\rm kin}}{H_{\rm end}}\)^{1/3} \simeq  cH_{\rm end}^{-1}\(\frac{T_{\rm end}}{T_0}\)\, ,
\end{eqnarray}
where ``MD'', ``RD'' and ``kin'' designate matter, radiation and
kinetic energy dominated epochs, $H_0$, $\Om_{\m 0}$ and $\Om_{\r
0}$ are the present values of the Hubble parameter, dimensionless
density parameters corresponding to matter energy density  and
radiation energy density, and $T_{\rm end}$ and $H_{\rm end}$ denote
the temperature and Hubble parameter at the end of inflation, respectively. We note that in Eq.(\ref{rom}),
we again ignore the insignificant logarithmic correction.

\section{Instant particle production and preheating}
\label{Instant Preheating}

We hereby emphasize once again that the gravitational particle production is quite
inefficient for circumventing the problem due to relic gravitational waves, one therefore should look for an alternative mechanism. One
of the alternatives suitable to the class of models similar to the
one we are going to discuss in the next section, is based upon instant particle production
dubbed instant preheating which proceeds as follows. We assume that
the scalar field, $\phi$, interacts with some other scalar field,
$\chi$, which interacts with the Fermion field, $\psi$,

\begin{equation}
\label{lag}
\mathcal{L}_{\rm int}=-\frac{1}{2} g^2 \phi^2 \chi^2-h \bar{\psi}
\psi \chi \,,
\end{equation}
where {\it g} and {\it h} denote the coupling constants, assumed to
be positive for convenience and $g,h<1$ for perturbation treatment
to be applicable. In this case, $\chi$ does not have a bare mass.
 However, the $\chi$ field has an
effective mass which grows with $\phi$ as $m_{\chi}=g |\phi|$. The
Lagrangian is specially design to give rise to this feature.
 As inflation ends, the field $\phi$
enters the kinetic regime as the potential is very steep in the
post-inflationary regime. Consequently, the field $\phi$  rolls down
its potential fast soon after inflation ends.
Since, mass of $\chi$ depends upon $\phi$, we can shift the field as $\phi\to \phi'=\phi-\phi_{\rm end}$ such that the effective mass of $\chi$ vanishes at the end of inflation. However, the Lagrangian (\ref{lag}) does not obey shift symmetry, we thus need to assume an enhanced symmetry in the Lagrangian which can be achieved by adding a suitable counter term to (\ref{lag}).
It is important to
check how $m_{\chi}$ changes with time around the transition from
inflation to kinetic phase. The non-adiabatic change in $m_{\chi}$ is
crucial for particle production.
Indeed, the production of $\chi$ particles take place
when the adiabaticity condition is violated  ,{\it i.e.},
\begin{eqnarray}
|\dot{m_{\chi}}| \gtrsim m_{\chi}^2 \Rightarrow |\dot{\phi}| \gtrsim
g \phi^2 \,.
\end{eqnarray}
The above condition is satisfied if
\begin{equation}
\label{ppoccur}
|\phi| \lesssim |\phi_p|=\sqrt{\frac{\dot{\phi}_{\rm
end}}{g}} \,.
\end{equation}
The equation of state parameter for the inflaton field is given by
\begin{equation}
\omega_{\phi}=\frac{\dot{\phi}^2-2V}{\dot{\phi}^2+2V}.
\end{equation}
Now using the fact that inflation ends when $\omega_{\phi}$ increases to -1/3, we find that
\begin{equation}
\dot{\phi}_{\rm end}=\sqrt{V_{\rm end}} \,.
\end{equation}
Considering the fact that $\phi_p \lesssim \Mpl$,
Eq.~(\ref{ppoccur}) gives us the bound on the coupling {\it g}
\begin{equation}
\frac{\dot{\phi}_{\rm end}}{g} \lesssim \Mpl^2 \rightarrow g \gg
\frac{1}{\Mpl^2} \sqrt{V_{\rm end}} \,.
\end{equation}
In addition, the production time of $\chi$ particles is given by
\begin{equation}
\delta t_p \sim \frac{|\phi|}{\dot{\phi}}= (g \dot{\phi}_{\rm
end})^{-1/2} .
\end{equation} 
The uncertainty relation allows us to obtain the momentum, $k_p \simeq (\delta p)^{-1} \simeq \sqrt{g
\dot{\phi}_{\rm end}}$.  Following the
Ref.~\cite{Felder:1999pv,Kofman:1997yn}, the occupation number
of $\chi$ is given by
\begin{equation}
n_k \sim \exp \(-\frac{\pi k^2}{k_p^2} \) \,.
\end{equation}
which allows us to estimate the number density of $\chi$ particles
\begin{equation}
N_{\chi}=\frac{1}{(2 \pi)^3} \int_0^{\infty} n_k d^3 \k = \frac{(g
\dot{\phi}_{\rm end})^{3/2}}{(2 \pi)^3} \, ,
\end{equation}
and their total energy density
\begin{equation}
\rho_{\chi}=N_{\chi} m_{\chi}=\frac{g^2 V_{\rm end}}{8 \pi^3} \,.
\end{equation}
Assuming that at the end of inflation, the produced energy is thermalized
instantaneously, we obtain
\begin{equation}
\(\frac{\rho_{\phi}}{\rho_r} \)_{\rm end} \simeq \frac{12 \pi^3}{g^2} \,.
\end{equation}
The above equation combined with Eq.~(\ref{eq:rhog_r_bound}) gives the lower bound on the coupling $g$
\begin{equation}
g \gtrsim 9.78 r^{1/2} \times 10^{-3}\,.
\end{equation}

\begin{figure}[h]
\centering
\includegraphics[scale=.7]{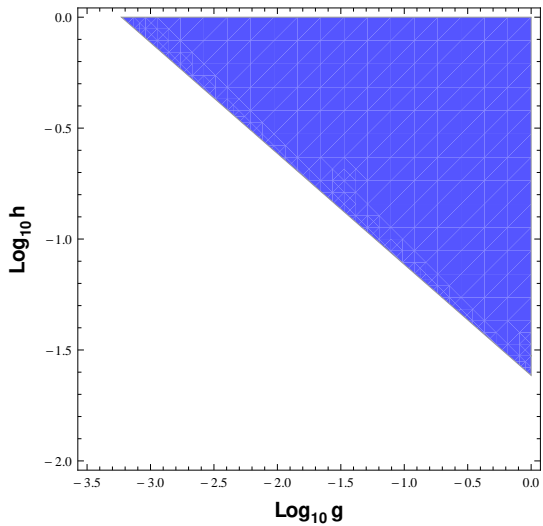}
\caption{Figure shows the parameter space of ($g,h$), while the
shaded region gives the allowed values of the parameters for efficient
preheating to occur. We have used $r=10^{-3}$ to plot this figure.} \label{fig:gh}
\end{figure}

Now, since
\begin{eqnarray}
\delta t_p H_{\rm end} &\simeq & \frac{1}{\Mpl \sqrt{2 g}} (V_{\rm end})^{1/4} < 9.29 \times 10^{-2}  \nn \\
\Rightarrow \delta t_p &\ll & H_{\rm end}^{-1} \,,
\end{eqnarray}
the expansion is negligible during the particle production.
It should be noted that since $\phi_p \lesssim 0.13 \Mpl$,
the production of particles commences immediately after the inflation ends. In the post-inflationary regime, the scalar field $\phi$ rolls down its potential faster and hence the effective mass $g \vert \phi \vert$ of the scalar field $\chi$ grows which facilitates the decay of  $\chi$ particles into fermions. The  decay rate given by
\begin{equation}
\Gamma_{\bar{\psi} \psi} = \frac{h^2 m_{\chi}}{8 \pi} = \frac{h^2}{8
\pi} g |\phi| \,.
\end{equation}
Around the end of inflation,  $\rho_{\phi} \propto 1/a^2$ and diminishes slower than $\rho_{\chi}$ with the expansion of universe\footnote{Immediately, after inflation ends ($\omega_{\phi}=-1/3$), $\rho_\phi\sim 1/a^2$, thereafter it redshifts faster and enters the kinetic regime. Since kinetic regime establishes fast after inflation, the estimates change insignificantly if we identify the end of inflation with the commencement of kinetic regime. }. We now should arrange the decay rate of $\chi$ into matter fields such that the decay takes place before the back reaction of $\chi$ on $\phi$ evolution  becomes important which implies that
\begin{equation}
\Gamma_{\bar{\psi} \psi} \gg H_{\rm end} \Rightarrow h^2 \gtrsim 8
\pi \frac{H_{\rm end}}{g |\phi|} \,.
\end{equation}
For $\phi \lesssim \Mpl$, the above condition gives us the lower
bound on the coupling {\it h}, namely, $h \gtrsim 0.13 g^{-1/2} r^{1/4}$. In Fig.~(\ref{fig:gh}), we depict the
allowed values of $g$ and $h$ which shows that we have a wide
parameter space for an efficient preheating to occur.

\begin{figure}[h]
\centering
\includegraphics[scale=.75]{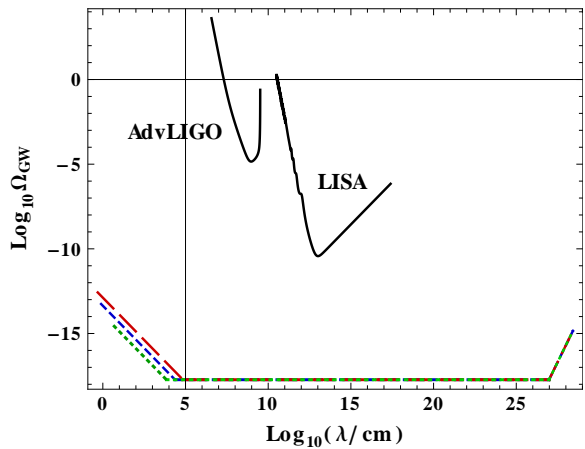}
\caption{Spectral energy density of the relic gravitational wave
background for different temperatures have been plotted. The blue (dashed) line corresponds to  $T_{\rm end}=6.29 \times 10^{13}$GeV obtained using COBE normalisation. While the red (long dashed) and green (dotted) lines have been plotted using the model we have considered in Section (\ref{sec:inflation}) for temperatures $T_{\rm end}=2.09 \times 10^{13}$GeV and $2.09 \times 10^{14}$GeV, respectively. The numerical values of model parameters are taken to be, $n=6$, $r=10^{-3}$ $\&$ $\lambda=8.08 \times 10^{-6}$. Also, we have considered $\mathcal{N}=65$ (behavior does not seem to
change significantly for the variation of $\mathcal{N}$), where
black solid curves represent the expected sensitivity curves of
Advanced LIGO and LISA.} \label{fig:RGW}
\end{figure}

Fig.~(\ref{fig:RGW}) illustrates the spectral energy density
($\Om_{\rm GW}$) of the relic gravitational wave background for the temperature $T_{\rm {end}}=6.29 \times 10^{13}$GeV which corresponds to $g=1.96 \times 10^{-3}$ (for $r \simeq 10^{-3}$) of the coupling, along with the sensitivity curve of AdvLIGO \cite{LIGO,aLIGO} and LISA
\cite{LISA,LISA1}. Figure shows that spectrum is
blue at high frequencies.

\section{Successful model of quintessential inflation} \label{sec:inflation}
We shall now check whether we can realize the aforesaid requirement in a particular model of quintessential inflation. To this effect, we shall consider a model based upon the following
generalized exponential potential with two parameters $n $ $\&$
$\lambda$,

\begin{equation}
\label{pot}
 V(\phi)=V_0 e^{-\lambda \phi^n/\Mpl^n}, n>1
\end{equation}
Using the  expressions of slow-roll parameters,
\begin{eqnarray}
\label{eq:slow-roll}
\epsilon \equiv \frac{M^2_{\text{Pl}}}{2} \left( \frac{V_{\phi}}{V} \right)^2~, \qquad \eta \equiv M^2_{\text{Pl}} \frac{V_{\phi \phi}}{V}.
\end{eqnarray}
we have  the standard expressions for the scalar spectral index
($n_s$) and tensor-to-scalar ratio ($r$) as
\begin{eqnarray}
\label{eq:spectnsr}
n_s - 1 = -6\epsilon + 2\eta ~, \qquad r = 16\epsilon \,.
\end{eqnarray}
In flat FRW background, the slow-roll parameter, $\epsilon$, for the
generalized exponential  potential (\ref{pot}) is given by:
\begin{eqnarray}
\label{epsilon}
\epsilon = \frac{1}{2} n^2 \lambda^2 \left(\frac{\phi}{\Mpl} \right)^{2n-2}\\
\end{eqnarray}
which shows that slow-roll takes place for $\phi/\Mpl \ll 1$ whereas
potential is steep for large values of the field. Using the
condition for the end of inflation, namely, $\epsilon \lvert_{\phi=\phi_{\rm {end}}} = 1$ and the expression for the number of the e-folding
\begin{eqnarray}
\label{eq:efolding}
\mathcal{N} &=& \int^{t_{\rm {end}}}_{t} H dt^{\prime} = -\Mpl^{-2} \int^{\phi_{\rm {end}}}_{\phi} \frac{V(\phi^{\prime}) d\phi^{\prime}}{dV(\phi^{\prime})/ d \phi^{\prime}} \\
&=& \frac{1}{n\lambda (n-2)} \left[ \left( \frac{\phi}{\Mpl}\right)^{2-n} - \left( \frac{2}{n^2 \lambda^2}\right)^{\frac{2-n}{2n-2}} \right] \,.
\end{eqnarray}
we estimate the numerical value of $\phi$ at the commencement of
inflation,
\begin{eqnarray}
\label{eq:phirel}
\frac{\phi}{\Mpl}= \left[ n(n-2)\lambda \mathcal{N} + \left( \frac{2}{n^2 \lambda^2}\right)^{\frac{2-n}{2n-2}} \right]^{\frac{1}{2-n}} \,.
\end{eqnarray}
We can then eliminate $\phi$ from slow-roll parameter, $\epsilon$, in
favor of model parameters $n,\lambda$ and the number of e-folds by
inserting the above expression for $\phi$ as
\begin{equation}
\epsilon = \frac{1}{2} n^2 \lambda^2 \left[n \lambda (n-2) \mathcal{N}+ \(\frac{\sqrt{2}}{n \lambda} \)^{\frac{2-n}{n-1}} \right]^{\frac{2 (n-1)}{2-n}} \, .
\label{eq:epsilon}
\end{equation}

We shall use the relation (\ref{eq:epsilon}) in the subsequent
discussion for the estimation of $V_{\rm
{end}}$. Current observations,
namely, the Planck 2015 results,  impose
 constraints on the model parameters $n$ $\&$ $\lambda$. Indeed, the
 predictions of the model are within $2 \sigma$ bound provided that
 $n\gtrsim 5$ and $\lambda\lesssim 10^{-4}$, see Refs \cite{Geng:2015fla,Geng:2017mic}.

The radiation energy  produced due to the production of particle at the end of inflation \cite{Ford:1986sy,Spokoiny:1993kt},
\begin{equation}
\rho_{\rm {r,end}} \simeq 0.01 g_p H_{\rm end}^4
\end{equation}
can now be calculated exactly.
 Choosing,  $r=10^{-3}$, $\mathcal{N}=65$, $n=6$ and making use of Eqs.~(\ref{eq:spectnsr}) and
(\ref{eq:epsilon}), we find that
$\lambda=8.08 \times 10^{-6}$. In addition, using COBE
normalization, we can estimate  $V_0=3.4 \times 10^{-8} r \Mpl^4$ and $V_{\rm end}=5.39 \times 10^{-9} r \Mpl^4$ so that
\begin{eqnarray}
\label{models}
\left(\frac{\rho_{\phi}}{\rho_{r}}\right)_{\rm {end}} \simeq \frac{\rho_{\rm \phi,end}}{0.01 g_p H_{\rm end}^4} \simeq \frac{1.11 \times 10^{11} g_p^{-1}}{r}.
\end{eqnarray}
Clearly, the radiation energy density is very low in this case, only one part in $10^{12}$ (assuming $g_p \simeq 100$ and $r \simeq 10^{-3}$) as compared to field energy
density at the end of inflation and violates the constraint on $\left(\rho_{\phi}/ \rho_{r}\right)_{\rm {end}}$ put by nucleosynthesis (\ref{eq:rhog_r_bound}). Also note that for this model, the ratio $H_{\rm inf}/ H_{\rm end}$, for $r \simeq 10^{-3}$, is found to be 1.87, which implies that $H_{\rm inf}$ is very close to $H_{\rm end}$ for $r \ll 1$ as we have mentioned in section \ref{RGW}. This ratio is larger for larger value of $r$.
For the model under consideration, one can invoke instant preheating mechanism as explained in the previous section. In this case, the lower bound 
 on the temperature at the end of inflation can be calculated using Eq.~(\ref{eq:rhog_rhor}) together with the constraint on ratio, $(\rho_g/\rho_r)_{\rm {rh}} \lesssim 0.01$, due to nucleosynthesis as,
\begin{eqnarray}
&&\rho_{\rm {r,end}} \gtrsim \frac{9 V_0^2}{\Mpl^4}~ \exp
\left[-\lambda {\left(n \lambda (n-2) \mathcal{N}+
\(\frac{\sqrt{2}}{n \lambda} \)^{\frac{2-n}{n-1}}
\right)^{\frac{n}{2-n}}}+\(\frac{\sqrt{2}}{n \lambda}
\)^{\frac{n}{n-1}} \right]\, \nn \\
&&\Rightarrow T_{\rm {end}}=\rho_{\rm {r,end}}^{1/4} \gtrsim 4.9~r^{1/2} \times 10^{14} \text{GeV}
\end{eqnarray}
which is consistent with the model independent estimate obtained earlier. Now that we have a lower bound on $T_{\rm end}$, using Eq.~(\ref{hor_cross}), we can obtain a bound on the number of e-foldings in this model, namely, $\mathcal{N} \lesssim 66.4
$. Note that the number of e-foldings depend on the temperature at the end of inflation and would be even more than this for the temperature of radiation due to gravitational particle production.

Also, the lower bound on the coupling constant $h$, for $n=6$, is found to be $h \gtrsim 0.12 g^{-1/2} r^{1/4}$. Clearly, the model under consideration successfully implements the paradigm of quintessential inflation. Last but not least, we should emphasize that it is challenging to find a model that can implement all the requirements of the scenario listed in the introduction.


\section{Conclusions} \label{sec:conclusion}
In  scenario of quintessential inflation, field enters the steep region of the
potential soon after inflation ends, such that $\rho_{\phi}\propto
a^{-6}$ ${\it a ~la}$ the kinetic regime. The duration of this
phase, which generally follows inflation in the scenario of the
quintessential inflation, irrespective of the field potential,
depends upon the temperature at the end of inflation. A longer kinetic regime or
the lower value of $T_{\rm end}$ implies the enhancement of energy density in relic gravitational waves compared to field energy density and might challenge  nucleosynthesis constraint at the commencement of the
radiative regime. The temperature at the end of inflation can be estimated from $H_{\rm inf}$ whose value is fixed using COBE normalization. The estimate on the  temperature is found to be $T_{\rm {end}}\simeq 1.99 r^{1/2} \times 10^{15}$GeV. We certainly need a reheating mechanism other than the standard one which is not operative in case of quintessential inflation. The gravitational particle production is a quantum mechanical process of particle creation from vacuum. The leading contribution to energy density of created particles in this process comes from the epoch when the transition from acceleration to deceleration takes place, namely, from the end of inflation. Using the estimate for $H_{\rm end}\simeq H_{\rm inf}$, we estimated the temperature of  radiation created during this process.  The radiation temperature so estimated turns out to be less than $T_{\rm end}$ challenging the nucleosynthesis constraint.

In order to set the goal, we
implemented the instant preheating which is based upon the
assumption that $\phi$ interacts with an auxiliary scalar field
$\chi$ with the coupling $g$ which then interacts with matter field
with the coupling strength $h$. We have found a wide range in the
parameter space, $(g,h)$ which allows to obtain the desired
 temperature (see Fig.~(\ref{fig:gh})).

The generic feature of the paradigm of quintessential inflation
includes relic gravitational wave background with blue spectrum
produced during the transition from inflation to kinetic regime. In
Fig.~(\ref{fig:RGW}), we depict spectral energy density of relic
gravitational waves for the temperature at the end of inflation which we have estimated model independently as well as for the temperatures we have calculated from the model we have considered. We have also plotted the proposed sensitivity curves for advanced LIGO and LISA. As seen in
the figure, the blue spectrum appears at high frequencies which
clearly distinguishes the paradigm of quintessential inflation from
conventional inflation. 

We show that the aforesaid requirements of quintessential inflation, which are model independent, can be successfully met by a model based upon the inflaton
potential, $V\propto Exp(-\lambda \phi^n/\Mpl^n), (n>1)$. This potential has an interesting property, namely, its slope goes as
$\phi^{n-1}$ giving rise to slow-roll for small $\phi$
whereas it exhibits steep behavior at late stages for large values
of the field. 
As demonstrated in Refs. \cite{Geng:2015fla,Geng:2017mic}, the model under consideration leads to a
viable post-inflationary dynamics. In particular, scaling solution
is an attractor of the dynamics, namely, an approximate scaling solution,  despite $n>1$ in (\ref{pot}). In this case, the
late-time exit from scaling regime to cosmic acceleration may be
successfully realized by invoking non-minimal coupling of the field
with massive neutrino matter \cite{Geng:2015fla,Geng:2017mic}. The present scenario, therefore,
is of great interest; it provides with a successful unification of inflation and late-time cosmic
acceleration.

\section*{ACKNOWLEDGMENTS}
We thank Md.~Wali Hossain, C.~Q.~Geng, C.~C.~Lee and E.~N.~Saridakis for useful discussions. S. Ahmad acknowledges DST, Govt. of India for financial support
 through Inspire Fellowship (DST/INSPIRE FELLOWSHIP/2012/614). MS
 thanks the Eurasian  International
Center for Theoretical Physics, Eurasian National University, Astana
for hospitality where the work was initiated. MS also thanks Maulana Azad National Urdu
University, Gachibowli, Hyderabad where part of the work was completed.

\end{document}